# *In silico* trials of acute ischemic stroke: predicting the total potential for improvement to patient functional outcomes

Claire M. Miller[1*,2], Raymond Padmos[1*], Praneeta Konduri[3,4], Tamás István Józsa[5,6], Yidan Xue[5*,7], Nerea Arrarte Terreros[3,4,8], Max van der Kolk[1*], Stephen Payne[5,9], Henk Marquering[3,4], Charles Majoie[4], Alfons Hoekstra[1#]

[1]Computational Science Laboratory, Informatics Institute, Faculty of Science, University of Amsterdam, Amsterdam, the Netherlands.
[2]Auckland Bioengineering Institute, University of Auckland, Auckland, New Zealand.
[3]Department of Biomedical Engineering and Physics, Amsterdam UMC, location University of Amsterdam, Amsterdam, the Netherlands.
[4]Department of Radiology and Nuclear Medicine, Amsterdam UMC, location University of Amsterdam, Amsterdam, the Netherlands.
[5]Institute of Biomedical Engineering, Department of Engineering Science, University of Oxford, Oxford, UK.
[6]Centre for Computational Engineering Sciences, Cranfield University, Cranfield, MK43 0AL, UK.
[7]School of Health Sciences, Faculty of Biology, Medicine and Health, The University of Manchester, Manchester, UK.
[8]Digital Health and Biomedical Technologies, Vicomtech, San Sebastian, Spain.
[9]Institute of Applied Mechanics, National Taiwan University, Taiwan.

## Abstract

This study uses *in silico* trials (ISTs) to quantify the potential for benefit due to improved recanalisation outcomes and shorter time to treatment for acute ischaemic stroke (AIS) patients. We use an IST framework to run trials on cohorts of virtual patients with early and late treatment after stroke onset, and with successful (full) and unsuccessful (no) recanalisation outcomes. Using a virtual population of AIS patients, and *in silico* models of blood flow, perfusion, and tissue death, we predict the functional independence of each patient at 90 days using the modified Rankin Scale (mRS).

Results predict 57% of the virtual population achieve functional independence with full recanalisation and a treatment time of 4 hours or less, compared to 29% with no recanalisation and more than 4 hours to treatment. Successful recanalisation was more beneficial than faster treatment: the best-case common odds ratio (improved mRS) due to recanalisation was 2.7 compared to 1.6 for early treatment.

This study provides a proof-of-concept for a novel use-case of ISTs: quantifying the maximum potential for improvement to patient outcomes. This would be useful during early stages of therapy development, to determine the target populations and therapy goal with the greatest potential for population improvements.

# 1 Introduction

Acute ischaemic stroke (AIS) occurs due to an occlusion, by a thrombus, of an artery supplying blood to the brain. The occlusion reduces blood supply to regions of the brain, leading to brain tissue death and a resulting loss of functional independence or, in some cases, death. The current standard of care for treatment of AIS is to restore the blood flow through lysis of the occluding thrombus using intravenous thrombolysis (IVT) followed by surgical removal through endovascular thrombectomy (EVT). The goal of treatment is recanalisation of the primary occlusive lesion and reperfusion of the downstream brain tissue, before tissue is irreversibly lost, and, as a result, improved functional independence for the patient [1–6].

The two main approaches to improving patient functional outcome in AIS are to improve the recanalisation success rate, and to reduce the time between occlusion onset and treatment. In the past decade, improvements to recanalisation rates through the standardised use of EVT, compared to IVT alone, have been substantial [3–7]. Improvements in workflow efficiency have, in some countries, shown improvement with the use of mobile stroke units (MSUs), with an associated improvement to patient outcomes [8–10]. These units, as an alternative to standard emergency services, can perform mobile computer tomographic scans and IVT.

Though the improvements to treatment efficacy and workflow efficiency have been extensive, they are also costly and slow to trial. Additionally, despite an increase in both reperfusion rates and time to reperfusion, over half of AIS patients still do not achieve functional independence [11]. Further development of therapies and workflow efficiencies are likely to require significant investment. Given this, it is useful to decouple the potential for further improvement to patient outcomes due to improvements in treatment outcome compared with workflow efficiency. This would allow for more streamlined investment in clinical trials. Such a decoupling could be performed effectively using *in silico* trials (ISTs).

ISTs are a recent concept in which computer simulation is used to test a proposed therapy on a virtual population of patients, either during product development or efficacy evaluation [12]. Previous examples of IST use include pre-clinical trial analysis of novel therapies and vaccines [13–15]; and post-clinical trial analysis to extend trial results [16], or to interrogate unexpected or inconclusive results [17,18]. ISTs have the ability to optimise dosage/application and the trial population of interest before any in-person administration begins, therefore reducing both the risk to the patient and the risk of trial failure.

To date, the majority of IST investigations have been proof-of-concept, and have not contributed, to our knowledge, to any clinical or regulatory decision-making in AIS therapy. One example of an IST framework that has been approved for use by the FDA is the UVA/Padova Type 1 Diabetes Simulator [19]. Work towards the goal of using *in silico* approaches as a standard component of development of new therapies is ongoing. The ASME V&V40 standard for credibility assessment [20] and the Avicenna Roadmap, developed by experts in the field [12], are examples of efforts to provide standards and guidelines for incorporating *in silico* approaches in therapeutic development.

In this paper, we demonstrate the use of an IST to determine the potential best-case outcomes for AIS patients with improvements to workflow times and recanalisation outcomes. Successful recanalisation in the model is the equivalent of an arterial occlusive lesion (AOL) recanalisation score of 3. We present a credibility study for our IST, comparing the patient outcomes for successful and unsuccessful recanalisation, to results from a real-life AIS clinical trial. We then run our IST to compare the potential for improvements in patient outcomes with increased recanalisation rates and improved workflow times. We define a best- and worst-case scenario for workflow times (early or delayed) and recanalisation outcome (successful or unsuccessful). We compare the difference in functional outcomes (at 90 days after AIS onset) for each of these scenarios to evaluate the maximum potential for improvement for each strategy.

## 2 Methodology

### 2.1 IST framework

We have presented our IST framework, 'des-ist', previously in Miller et al. [21] (methodology) and van der Kolk et al. [22] (software implementation). The des-ist framework uses an event-based modelling approach to run multi-model trials on a population of virtual patients, generated using a statistical patient generation model. Its use has previously been demonstrated for ISTs of EVT treatment in AIS [23]. In this study, we use the framework to run mechanistic models of arterial blood flow, blood perfusion through the brain, and tissue death on a cohort of virtual patients to predict patient infarct (dead tissue) volume. We then employ a statistical prediction tool to estimate each patient's functional independence after 90 days. The workflow is shown in Fig. 1.

### 2.2 IST Models

The models used in this study have been detailed in previous publications [21,23–29], so here we only provide a brief overview. The first model in the pipeline is the virtual patient generation model, which has previously been used in our IST framework [21,23]. Our virtual population model is based on data of patients enrolled in the MR CLEAN Registry: an observational, multi-centre study from 16 intervention hospitals in the Netherlands [3]. In brief, the virtual patient generation model generates a cohort of virtual patients using a statistical modelling approach, given some exclusion/inclusion criteria for the cohort. The attributes of these virtual patients are relevant and prognostic clinical characteristics [30]. The use of a statistical model enables us to generate any number of patients with combinations of characteristics that would be likely to appear in the clinic.

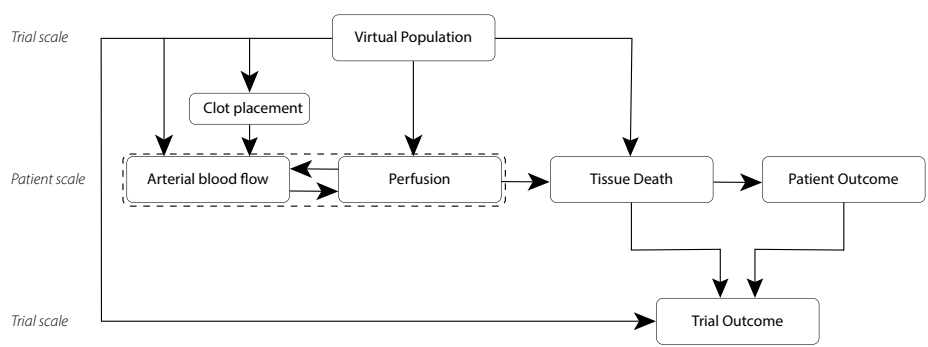

**Figure 1**: *Flowchart of the models included in the IST framework (des-ist). Data flow between models is indicated by the arrows. Dashed line indicates that the two way coupled models (arterial blood flow and perfusion) are incorporated as a single module in the framework. The coupled arterial blood flow and perfusion models are solved for three time points: once before the occlusion event; once after the occlusion occurs; and finally at the treatment time point (with or without an occlusion).*

The infarct volume (volume of dead tissue) and location at 24 hours after treatment is predicted for each patient using mechanistic models for blood flow, brain perfusion, and tissue death, as shown in Fig. 1. Detailed descriptions of each of these models can be found in [24–29], and their implementation in the framework can be found in [21,31]. In brief, using patient characteristics generated from the virtual patient model, these models predict the associated arterial flow to the brain [24] and resulting perfusion of blood through the brain tissue [25,26]. This is done in the non-occluded state (prior to onset), and then again with the occlusion inserted in the arterial network, and for a final time after treatment. Based on the change in tissue perfusion, and the amount of time until recanalisation (assuming full reperfusion), the region of hypoxic tissue at 24 hours is estimated [21,28].

The clinical score we predict in this study is the mRS (at 90 days), which scores the patient's level of independence on a scale from 0 (complete independence) to 6 (death). Functional independence is defined as an mRS score of 0 to 2 (no symptoms to slight disability. The patient outcome model predicts the patient's mRS score using the predicted infarct volume and their clinical characteristics. More detail on this model can be found in [32,33].

The final module is 'Trial Outcome', which generates a document that summarises the results of the IST, including the virtual population characteristics and functional outcomes, in an easy-to-read manner, suitable for clinical evaluation.

## 2.3 In silico trials design

For this study, we use an IST to investigate the potential for improvement in patient outcomes due to improvements in time to treatment and recanalisation outcome. We define the treatment time as the duration from stroke onset to start of EVT procedure. We generate two virtual populations: one treated within a short time window (early) and one treated after this window (delayed). An early time to treatment is defined as 4 hours or less, commensurate with the median time from onset to EVT in a meta-analysis of five clinical trials by Saver et al. [34]. A delayed time to treatment is defined as 4 hours or more. We note most patients in the MR CLEAN Registry, on which the virtual population model was built, were treated within 6 hours. On each of the generated virtual populations we run the trial pipeline twice: once where we assume all patients had a successful treatment outcome (complete recanalisation of the primary occlusive lesion) and once where we assume all patients had an unsuccessful treatment outcome (no recanalisation). This allows us to compare the effect of treatment and time to treatment independently, as well as the interaction of the two. We denote the two treatment outcome scenarios as successful and unsuccessful, and the two time to treatment scenarios as early and delayed. Each virtual population consists of 256 patients, to balance computational time and sufficient patient numbers to obtain a representative distribution. This patient count is also chosen to be similar to the number of patients allocated to each arm of the MR CLEAN Trial [7,35], the dataset we use for our credibility study, as described below in Section 2.4.

For this study we considered patients with occlusions of the M1 section of the middle cerebral artery, the most common large vessel occlusion location for AIS (64% and 58% of patients in MR CLEAN Trial and Registry respectively [3,7]). The majority of model validation has previously been performed using data from M1 occlusion patients, so we consider the framework to be most credible for these patients.

Before we present the results from the trial of interest for this study, we will present results from a credibility study on the IST. This study compares the IST outcome to the MR CLEAN Trial dataset [7,35], which we also split into successful and unsuccessful recanalisation outcome, as detailed in Section 2.4. The virtual population for the credibility study spans the full time frame for treatment, as used in the MR CLEAN trial, but only considers patients with M1 occlusions, to most closely align with the scope of the IST.

## 2.4 Clinical data

We use clinical data from the MR CLEAN Trial [7,35] to perform a credibility study for our IST. The MR CLEAN trial was a clinical trial, in the Netherlands, evaluating the use of intraarterial treatment for AIS. There are multiple clinical indicators for recanalisation in a clinical trial. We use the arterial occlusive lesion (AOL) recanalisation score, with successful recanalisation defined as an AOL of 3 and unsuccessful recanalisation outcome defined as AOL< 3. The AOL score is a clinical score that grades the level of distal flow and recanalisation of a lesion. We only included patients with M1 occlusions, and patients with no reported AOL were excluded from the analysis. This resulted in 129 patients in total, of which 66 had successful recanalisation.

# 3 Results

## 3.1 Credibility study

We first assess the credibility of our IST framework for the proposed study. There are two outcomes of interest for the IST in this study: distribution of infarct volume, and the common odds ratio for the change in the distribution of mRS. Consequently, the context of use, defined based on the V&V40 Standard [20], is to run the IST for large cohorts of patients for both successful and unsuccessful treatment scenarios to predict the difference in the distribution of infarct volume and mRS. In this section we assess the credibility of the trial for this context of use by comparing the infarct volume and mRS distributions from the MR CLEAN trial, split by successful or unsuccessful recanalisation, to the IST predictions for the population.

### 3.1.1 Infarct volume

The first outcome of interest is the distribution of infarct volumes for the successful and unsuccessful recanalisation outcome scenarios. Infarct volume is the volume of dead tissue, which we predict at 24 hours after treatment. A comparison of the trial results and the clinical results is shown in Fig. 2(a,b) and summary statistics are given in Table 1a. The IST estimates the mean to within 15 mL of the clinical data, for both recanalisation outcomes, and we observe both a reduction in median value and range when recanalisation is achieved. However, despite this close prediction, the IST is lower in the successful recanalisation arm and higher in the unsuccessful result compared to the clinical data. As a result, the reduction in mean infarct volume due to successful recanalisation is 36% and 57% for the clinical and IST results respectively. Additionally, the third quartile infarct volume is much higher in the IST for unsuccessful recanalisation patients and, to a lesser extent, lower in the successful recanalisation arm. Consequently, though it provides a close prediction of the mean for each arm individually, the IST is over-predicting the effect of successful recanalisation on infarct volume.

We can use a bootstrapping approach, resampling the clinical data ($10^5$ bootstrapped trials, 256 patients per trial), to estimate the distribution of the mean infarct volume of a sample population. We estimate a 95% confidence interval (CI) on the mean of (71 mL, 88 mL) in the unsuccessful recanalisation case, and (44 mL, 58 mL) in the successful recanalisation case. This further reinforces the observation that the IST overestimates the effect of successful recanalisation on population infarct volume.

There are several potential explanations for the over-prediction in the unsuccessful recanalisation case. Firstly, our unsuccessful recanalisation category for the clinical data includes patients who had a partial recanalisation with some distal flow (i.e. an AOL score of 2), whereas the IST only considers a binary outcome of complete thrombus or no thrombus. We also only included patients with lesion volume recorded at 24 hours, which excludes any patients who did not survive to 24 hours. Both of these limitations may bias the clinical data towards lower infarct volumes in the unsuccessful recanalisation case. Finally, the population characteristics (Table S1) for the IST has a smaller proportion of patients (17.2% compared to 36.5%) with a collateral score of 3 (high collateral flow to the occluded region). The difference in collateral flow between individuals is implemented through the boundary conditions of the model, as detailed in Miller et al. [21]. A lower proportion of patients with high collateral score would increase the mean infarct volume of the population, particularly with unsuccessful recanalisation. With respect to the under-prediction in the successful recanalisation arm, the IST also has a shorter duration between onset and EVT, and, compounding this, we do not incorporate the time of the EVT treatment itself in our model–the median duration of which was 63 minutes in MR CLEAN Registry (first cohort) [11]. The associated reduced time to treatment would have an effect of reducing the infarct volume with successful recanalisation.

### 3.1.2 Functional outcome

The clinical outcome of interest in clinical trials of AIS is the functional outcome, or level of independence of the patient. In this trial, we use the mRS at 90 days as the functional outcome metric. Specifically, we are interested in understanding the improvement to the distribution of mRS seen with improvement to treatment. The change in this distribution is commonly measured using the common odds ratio [7], which, under the standard mRS order, summarises the odds of a patient obtaining a higher mRS score due to the change in care. To make this more intuitive, here we reverse the mRS scores to determine the ratio in the direction of functional improvement (i.e. common odds ratio greater than one indicates a higher probability of a lower mRS score). We also report the proportion of patients who attain functional independence, defined as an mRS of 0 to 2.

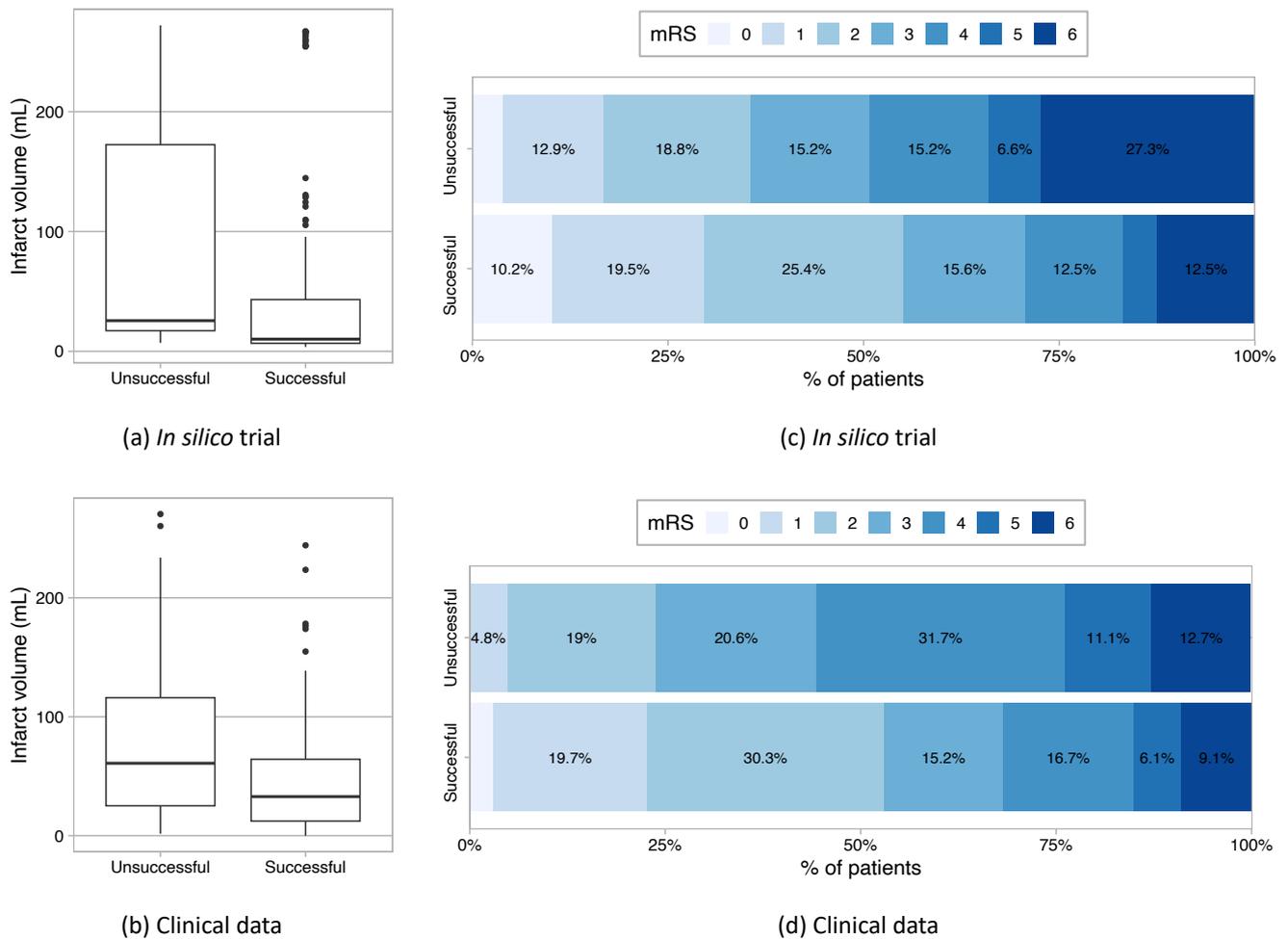

*Figure 2*: Distributions for the main outcomes of interest for the credibility study. Results from the IST and clinical data are compared between patients with successful (AOL of 3 for the clinical data) and unsuccessful recanalisation. (a-b) Box plots of the infarct volume at 24 hours for the IST (a) compared to the clinical data (b). (c-d) mRS distribution for IST (c) compared to the clinical data (d).

The distribution of mRS scores can be seen in Fig. 2(c,d) for both the IST and the clinical data. Figure 2 shows the IST has a notably higher proportion of unsuccessful recanalisation patients with an mRS of 6 and a lower proportion of unsuccessful recanalisation patients with an mRS of 3. For the successful recanalisation arm, the IST has a notably higher proportion of patients with an mRS of 0. As seen for infarct volume in Section 3.1.1, the IST is overpredicting the effect of recanalisation on functional outcome.

These differences are less significant when we calculate the summary metrics of interest, the log odds shift and proportion of population with functional independence, given in Table 1. Using the binary category of functional independence (mRS 0–2) or not (mRS 3–6), the IST is a close match in the successful recanalisation case, but is higher in the unsuccessful recanalisation case. Opposite to what we observed in the infarct volumes, this indicates the IST is under-predicting the effect of recanalisation on the proportion of patients who achieve functional independence. The log odds shift of the IST, given in Table 1(b), provides high confidence in the credibility of the IST to predict improvements to the population mRS distribution.

## 3.2 *In silico* trial results

### 3.2.1 Population characteristics
The characteristics of the two virtual populations, early and delayed treatment, are given in Supplementary Table S3. There is no significant difference in the population characteristics of the two trial arms, except in the workflow times as required for the trial.

### 3.2.2 Infarct volume
The first outcome of interest is the infarct volume distribution. The predicted impact of earlier treatment and improved treatment outcomes on infarct volume is shown in Fig. 3(a). These results are summarised in Table 2(a). The relative reduction in average infarct volume due to recanalisation ranged from 40% for delayed treatment patients to 60% for early treatment

|  | Num. | Infarct volume (mL) | Func. Ind. (%) |
| --- | --- | --- | --- |
| Unsuccessful | | | |
| *In silico* trial | 256 | 90.6 (17.2–172.5) | 36 |
| Clinical data | 63 | 79.5 (25.1–116.0) | 24 |
| Successful | | | |
| *In silico* trial | 256 | 39.0 ( 6.7–43.2) | 55 |
| Clinical data | 66 | 50.7 (12.3–64.2) | 53 |

|  | Estimate | 95% CI |
| --- | --- | --- |
| *In silico* trial | 2.3 | (1.7, 3.2) |
| Clinical data | 2.9 | (1.6, 5.6) |

(a) Clinical outcomes

(b) Common odds

*Table 1*: Summary metrics for the credibility IST compared to clinical data. (a) Number of patients (Num.), infarct volume (at 24 hours), and percentage that achieve functional independence (Func. Ind.) for patients with successful (AOL 3 in clinical data) and unsuccessful recanalisation outcome. Infarct volume results given as 'mean (inter-quartile range)'. (b) Common odds ratio due to successful recanalisation. CI: Confidence interval.

patients. The reduction due to decreased time to treatment varied between -8% (8% increase) for unsuccessful recanalisation patients and 29% for successfully recanalised patients. The overall potential for reduction in mean infarct volume from the worst-case (delayed, unsuccessful recanalisation) to the best-case (early, successful recanalisation) treatment scenario is 57%.

The 8% increase seen here with decreased time to treatment can be explained by differences in the population characteristics between the early and delayed treatment arms of the trial. In the unsuccessful recanalisation case, for the same virtual population, the infarct volume predicted by the model would be identical regardless of the time to treatment. However, in this study, the virtual populations are different for the two trial arms. In particular, the delayed treatment population had 42.1% of patients with bad collaterals (0 or 1), compared to 46.0% for the early treatment trial (Supplementary Table S3). Smaller effects could also be due to higher systolic blood pressure in the early treatment arm, which would also have an effect of increasing infarct volume, however this is expected to be fairly insignificant compared to the difference in collateral scores distribution.

We see a more marked improvement if, rather than the mean volume, we consider the third quartile (Q3) values. The interpretation is that 75% of patients achieving an infarct volume less than or equal to the Q3 value. Here, we see a relative reduction of 51% and 76% due to recanalisation for the delayed and early treatment scenarios respectively; and -5% and 50% due to a reduced time to treatment for the unsuccessful and successful recanalisation scenarios respectively. The overall potential for reduction in Q3 volume between our worst-case and best-case scenarios is 75%. However, it is important to note that, in the credibility study, we observed an over-prediction of the Q3 infarct volume in the IST results for the unsuccessful recanalisation population.

In Fig. 3(a), we also obserev a tri-modal response of the infarct volume distribution in the unsuccessful recanalisation arms. For patients who do not have successful recanalisation, the only significant difference in infarct volume between patients after 24 hours, according to the mechanics built into the model, will be the level of collateral flow to the occluded region. The amount

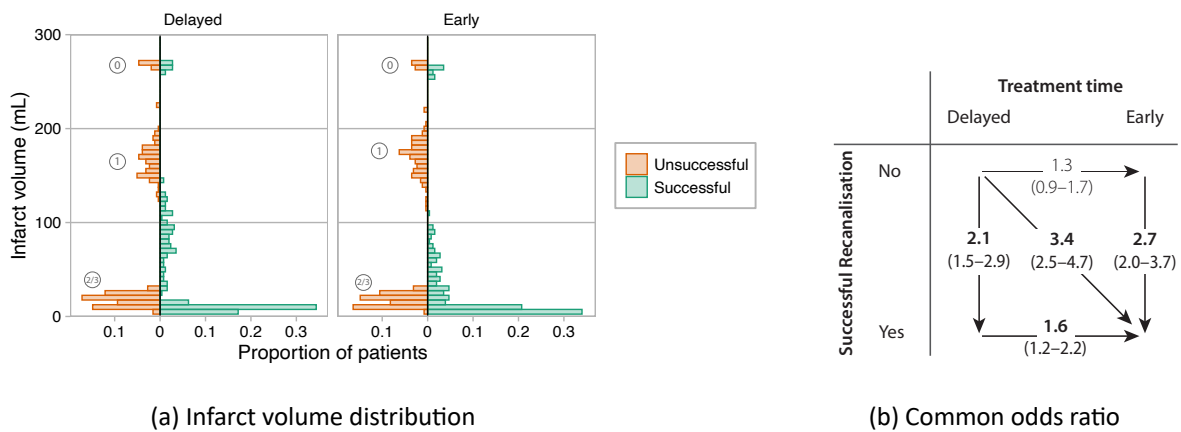

(a) Infarct volume distribution

(b) Common odds ratio

*Figure 3*: Outcomes of interest for the IST. (a) Infarct volume distributions for each of the four trial arms: early and delayed treatment, with both unsuccessful and successful recanalisation. The annotations (0), (1), (2/3) indicate the collateral scores associated with the different modes seen in the distribution. (b) Common odds ratio between the different arms of the trial, showing the improvement in functional outcome (mRS) due to both treatment outcome and decreased time to treatment. Ranges in brackets give the 95% confidence interval.

|             | Delayed           | Early             |
| ----------- | ----------------- | ----------------- |
| Unsuccessful| 87.3 (17.1–166.6) | 93.9 (17.4–174.4) |
| Successful  | 52.5 ( 8.9– 81.9) | 37.4 ( 6.3– 41.4) |

(a) Infarct Volume (mL)

|             | Delayed | Early |
| ----------- | ------- | ----- |
| Unsuccessful| 28.5    | 35.2  |
| Successful  | 46.9    | 57.0  |

(b) Functional independence (%)

*Table 2*: Outcomes of interest for the different trial arms: delayed/early treatment and unsuccessful/successful recanalisation outcome scenarios. (a) Infarct volume, given as 'mean (inter-quartile range)'; and (b) percentage of patients with functional independence (mRS ≤ 2).

of collateral flow for a patient is scored in the clinic using a grading of 0 to 3 (bad to good collaterals). Consequently, the three modes seen in Fig. 3(a) are collateral scores of 0, 1, and 2/3, in order of decreasing mode, as shown on the figure. When recanalisation is successful this effect is less apparent, as volume also depends on the time of the recanalisation in this case.

### 3.2.3 Functional outcome

The second outcome of interest for the IST is the functional outcome, defined by the mRS distribution. For simplicity of comparison between the four arms of the trial we only give the proportion of patients with functional independence here. The full mRS distributions for each trial arm can be found in Supplementary Fig. S1. Functional independence is defined as an mRS score of 0 to 2, and the proportions are given in Table 2(b). Results show the maximum potential improvement to patient outcome, from our worst-case scenario (delayed, unsuccessful) to our best-case scenario (early, successful), is functional independence for an additional 28.5% of patients. The majority of this is provided by recanalisation, with functional independence achieved in an additional 21.8% of patients with improved recanalisation alone. In comparison, the maximum potential for improvement due to a reduction in time to treatment alone is an additional 10.1% of patients. Given the infarct volume between the delayed and early treatment arms for the unsuccessful recanalisation case are similar, one might have expected that the functional dependence would also be very similar. The observed improvement to outcome with early treatment in the unsuccessful recanalisation case is a result of the patient outcome model, whose predictors include several patient characteristics in addition to infarct volume, including the time to treatment, to predict patient mRS scores [32]. Other studies have also shown that infarct volume is only one of several predictors of clinical outcome [36].

In addition to the absolute values for the mRS, the common odds ratios are shown in the diagram in Fig. 3(b). As in the credibility study, we reverse the mRS order, so a higher ratio indicates a shift towards better functional outcome (lower mRS score). The potential for improvement to the population distribution is shown to be very high with a ratio of 3.4. For comparison, the log odds score for improvement to population outcomes due to EVT has been estimated at 2.26 [5]. As above, the most significant potential for improvement is seen to be due to improved recanalisation rates, with a maximum potential log odds of 2.7 compared to a maximum of 1.6 due to improvements to treatment time alone. We also observe the lower bound of the 95% confidence interval for the common odds ratio under reduced time to treatment with no recanalisation is 0.9. This indicates there is a possibility that there is a small negative effect associated with faster treatment when recanalisation is not achieved, however more data would be required to confirm this.

We can also quantify the improvement due to a reduced time to treatment in a continuous manner using the common odds ratio for every 30 min delay to treatment, as done in Menon et al. [37]. Combining the two trial populations (early and delayed treatment), we get a common odds ratio of 0.94 (p=0.025, 95% CI: 0.892–0.992) per 30 min delay. Further detail on this can be found in Supplementary Fig. S2.

## 4 Discussion

### 4.1 Credibility of the IST

We assessed the credibility of the in silico trial (IST) for predicting the infarct volume, mRS distribution, and functional independence improvement due to the effect of recanalisation. The IST is determined to have high credibility for assessing the mean infarct volume under a specified treatment outcome. It is also shown to have a high level of credibility for assessing the proportion of patients who achieve functional independence with successful recanalisation outcome, and the log odds shift to a better outcome of the mRS distribution. This gives us high confidence in using our IST to estimate the potential improvements due to recanalisation and time to treatment on population outcomes, in particular through the common odds ratio as a metric to estimate mRS distribution improvements. However, the IST may over-predict improvements to infarct volume, and under-predict improvements to the proportion of patients who achieve functional independence, which should be taken into account when interpreting results.

## 4.2 Potential for improvement to patient outcomes

In this study, we investigate the maximum potential for improvement to patient outcomes under the best-case treatment scenario, in comparison to the worst-case treatment scenario. Our IST predicts the maximum potential for functional independence in AIS (M1 occlusion) patient outcomes, under our best-case treatment scenario, to be 57% of patients. This is equivalent to an additional 28.5% of patients achieving functional independence, or double the percentage, compared to our worst-case scenario. This potential for improvement is determined using the best-case scenario: all patients achieve recanalisation within 4 hours of AIS onset, compared to the worst-case scenario: no recanalisation and treatment after 4 hours. The predicted reduction in mean infarct volume from the worst- to best-case scenario is 57%, with an associated common odds ratio of 3.4 towards improved functional outcome. Recent studies from the MR CLEAN Registry dataset estimate current functional independence rates after AIS to be 42.6% (as of Nov. 2017), with successful reperfusion achieved in 65.7% of patients and a median time from onset to groin puncture of 180 minutes [11]. Our results indicate a potential for an additional 14.4% of patients to achieve functional independence with improved treatment, however it is important to note the scope of this registry is not the same as this study, as the MR CLEAN Registry data includes multiple occlusion locations.

As expected, the mean infarct volume is significantly reduced in the successful recanalisation arms compared to the unsuccessful recanalisation arms of our IST, with a potential for reduction of at least 40% and at most 60%. Due to this decrease in infarct volume, at a minimum an additional 18.4% of patients have the potential to achieve functional independence, with a maximum potential of an additional 21.8%. These improvements are associated with a common odds ratio, towards an improved mRS, of 2.1 and 2.7 respectively. This is comparable to the ratios observed clinically, due to the use of thrombectomy as a therapeutic strategy: 2.26, from a meta-analysis of five clinical trials [5].

We predict an upper bound on the potential mean infarct volume reduction due to a decreased time to treatment, from more than to less than 4 hours, to be 29% if all patients achieve recanalisation. Associated with this prediction, a maximum of 10.1% more patients achieve functional independence (successful recanalisation arm), with a common odds ratio of 1.6. In the context of a real clinical trial this would be considered a significant improvement—this is close to the improvement observed in the MR CLEAN trial [7]—however it is not as significant as the potential improvement predicted due to recanalisation.

We would expect little difference in the outcome for improved time to treatment in the case where the occlusion remains in the patient. Counter-intuitively, we see in Fig. 3(a) and Table 2(a) that the early treatment arm has a marginally higher infarct volume compared to the delayed treatment in the unsuccessful recanalisation case. As detailed in Section 3.2.2, this can be explained by variations in other population characteristics between the two trial arms. However, is not reflected in the populations mRS distribution summary metrics (Table 2(b) and Fig. 3(b)), due to the effect of earlier treatment in the statistical patient outcome model, as well as the effect of other clinical characteristics, such as age and NIHSS at baseline, in this statistical model.

We can compare our study results to previous clinical trial outcomes detailing patient outcomes according to time and treatment success. Results from the HERMES collaboration [34], a meta-analysis of several EVT trials, estimated the probability of functional independence to be 61% if the patient achieved substantial reperfusion at 4 hours. In comparison, we predicted a lower functional independence with successful recanalisation: 57% of patients treated within (rather than at) 4 hours. However, substantial reperfusion in HERMES is defined as modified treatment in cerebral infarction (mTICI) 2B/3, which is not a direct measure of recanalisation success. For example, in DEFUSE 3, a clinical trial for the use of EVT after 6 hours, only 48% of mTICI 0 to 2A were patients who were recanalized [38]. Additionally, the HERMES collaboration was not limited to M1 occlusion patients (70.5% M1 in HERMES).

An analysis of the ESCAPE Trial [4], a clinical trial on rapid use of EVT, by Menon et al. [37] found that, for every 30 min delay in time from onset to reperfusion, the odds ratio of functional independence was 0.912 (p=0.044). The ESCAPE trial only considered patients who achieved reperfusion with moderate-to-good circulation—this is equivalent to a collateral score of 2–3, and multiple occlusion locations (68% M1). In comparison, in the DEFUSE 3 trial, a one hour delay in onset to EVT arterial puncture time was associated with a 0.86 odds ratio shift in mRS [38]. However, this trial only included patients who presented at the clinic 6–16 hours after stroke onset. If we perform the same analysis on our IST results, using only good collateral patients, the time to treatment is not found to be significant (Supplementary Fig. S2). This is likely due to the slow growth rate of infarct volume under good collateral flow in our model. Using the full virtual population the odds ratio of functional independence for every 30-minute delay is 0.94 (p=0.025, 95% CI: 0.89–0.99, Supplementary Fig. S2).

## 4.3 Study and modelling limitations

As noted previously, our study is limited to AIS patients with an M1 occlusion. Additionally, the data used for both model development and the credibility study is from a cohort of patients from the Netherlands treated using a combination of IVT and EVT [3,7,35]. Consequently, the credibility study is only applicable to this population.

We do not consider current recanalisation success rates due to current standard of care (estimated at around 70%), or the potential value of improved time to treatment for patients outside of the standard 6-hour treatment window. For these patients, EVT is often not performed as its efficacy is less established in patients after 6 hours, though this is the subject of both recent and current clinical trials [39–41]. An improvement in time to treatment for late presentation patients would likely result in significant improvements in outcomes. Additionally, a different definition of 'early treatment' in this study may change the observed effect.

The virtual population generation model is limited to the clinical data that was collected during the clinical trial. For example, the data uses a categorical variable for collateral flow to the affected region. This is a coarse discretisation of what is likely a continuous variable and influences our infarct volume prediction. We also do not include the time spent administering treatment in the virtual population and, as a result, assume recanalisation occurs at the time of treatment, potentially biasing our results towards improved outcomes for recanalised patients. For example, in the MR CLEAN Registry the median duration of EVT was 54/63 minutes (first/second cohort) [11], and in the ESCAPE trial the median time from groin puncture until reperfusion was 30 minutes [4].

An event-based modelling approach is a useful method for building multi-model frameworks, however there are some limitations associated with this approach. Firstly, the method assumes that changes to the system are instantaneous in comparison to the time scales considered by the models. Though this seems reasonable for any change to blood flow conditions through the arterial system, it would be expected that the thrombus characteristics, and hence permeability, would change throughout the duration of the stroke, and hence affecting the blood flow and perfusion. Such changes could be due to increased compression of the clot over time, or the effect of both natural thrombosis and thrombolysis.

Another limitation of an event-based modelling system is that the coupling between models can only be one-way and can only occur at each event, unless fully coupled models are incorporated as a single model within the framework. For example, in this study, the (arterial) blood flow and perfusion are fully coupled and implemented as a single coupled model to solve for each event state. However, the tissue death model takes the perfusion state as an input but does not couple back to the perfusion model, so there is no effect of tissue death on perfusion. It is known that the process of tissue death likely affects the blood perfusion within the affected region and its surrounds. Detailed analysis of the limitations of these models have been discussed in previous publications [24–29].

Finally, we do not explicitly model any adverse events, such as haemorrhage or thrombus fragmentation. Though these effects are statistically accounted for in the patient outcome model, the model cannot predict any change in these events due to improved treatment techniques. These outcomes assume patients have been treated with the current standard of care: EVT, usually in combination with IVT (alteplase).

### 4.4 Summary

This study provided estimates for the maximum potential for functional independence with improved recanalisation outcomes and time to treatment for AIS patients with an M1 occlusion. It demonstrated the use of ISTs as a tool to quantify the maximum potential for improvement to patient outcomes based on improvements to current standards of care. This novel application of ISTs has the potential to assist investment decision making processes, and our study provides a proof-of-concept of how an IST could be used to support decision making of this type. The study also presented an approach to quantitatively assessing the credibility of an IST. Future work will include further development of the underlying models included within the IST framework and approaches for effective assessment of IST credibility.